\documentstyle[prb,aps,psfig,eqsecnum,multicol]{revtex}

\newcommand{\crossprod}{\times}
\newcommand{\be}{\begin{equation}}
\newcommand{\ee}{\end{equation}}
\newcommand{\barr}{\begin{eqnarray}}
\newcommand{\earr}{\end{eqnarray}}
\newcommand{\breakeq}{\nonumber \\ &&}

\begin{document}
\draft

\title{Theory of a magnetic microscope with nanometer resolution}

\author{Peter Johansson\cite{PJmail}} 
\address{Division of Solid State Theory,  Department of Physics,
   University of Lund, S\"olvegatan 14 A, S-223\,62 Lund, Sweden}
\address{and Department of Natural Sciences,  University of \"Orebro,  
          S-701\,82 \"Orebro, Sweden}
\author{S. Peter Apell\cite{PAmail}}
\address{Department of Applied Physics, Chalmers University of Technology and
G\"oteborg University, S-41296 G\"oteborg, Sweden  \\ and  
Donostia International Physics Center, San Sebastian, Spain} 
\author{D. R. Penn\cite{DPmail}} 
\address{Electron Physics Group,  National Institute of Standards and 
          Technology,  Gaithersburg, MD 20899, USA}  
\date{\today}
\maketitle
\begin{abstract}
We propose a theory for a type of apertureless scanning near field microscopy 
that is intended to allow the measurement of magnetism on a
nanometer length scale. A scanning probe, for example 
a scanning tunneling microscope (STM) tip, is
used to scan a magnetic substrate while a laser is focused on it.
The electric field between the tip and substrate is enhanced
in such a way that the circular polarization due to
the Kerr effect, which is normally of order 
0.1\%\, is increased by up to two orders of magnitude for the case of a Ag
or W tip and an Fe sample. Apart from this there is a large background of
circular polarization which is non-magnetic in origin. This circular
polarization is produced by light scattered from the STM tip and substrate. 
A detailed retarded calculation for this light-in-light-out
experiment is presented.
\end{abstract}


\begin{multicols}{2}

\section{Background}

A microscope that can measure magnetic structure with high resolution would 
be very desirable for practical reasons such as investigating read heads and 
media for magnetic recording and because magnetic nanostructures are currently 
of great interest. We propose a scanning magnetic microscope with a 
resolution of approximately 1 nm to 10 nm \cite{Resolvnote} depending
on the tip shape. 
This device is a form of apertureless  near-field scanning magneto-optical 
microscope in which light from a laser is focused at the tip of a scanning
tunneling microscope (STM) or atomic force microscope (AFM) or at
an isolated sphere on the sample surface 
and the 
circular polarization of the light scattered from the tip and sample is 
measured. This is analogous to attempts to use the circular polarization 
of the light emitted by a scanning tunneling microscope
to construct a magnetic microscope.\cite{Vazques,Majlis,Pierce,APJ}
It was found that the light-emission  scheme is not suitable 
for use as a magnetic microscope due to the small degree of circular 
polarization produced by a magnetic sample as well as the relatively low 
intensity of the emitted light.\cite{Pierce} The light intensity
is low despite the fact that the electric field at the tip of an 
STM can be strongly enhanced by interface plasmons localized between
the sharp tip apex and the substrate. Without the enhancement the
light intensity would be smaller by several orders of magnitude. Other
techniques to develop magnetic microscopes based on 
apertureless near-field scanning have resolution on the order of at 
least 100 nm.\cite{silva,smol,aigouy} 
Attempts to measure magnetic 
surface structure by means of the STM tunneling current\cite{Crstep}
have also been made and recently the magnetic structure of
an antiferromagnetic overlayer was mapped using this technique.\cite{Mnlay}

In the scheme we propose, the scattered laser light 
intensity is large compared to that produced in STM light emission 
experiments making detection far easier. The circular polarization 
of the scattered light has both a component that is independent of 
the state of magnetization of the sample and a component that 
depends on the magnetization. The latter effect is rather small
both when light is scattered from a magnetic surface as in the 
surface magneto-optic Kerr effect, and when light is 
emitted from an STM.  
However, when a tip is present and a laser provides incident light
that is linearly polarized
parallel to the surface (s polarized) and propagating in the direction
of magnetization, we find up to
two orders of magnitude increase in the degree of circular polarization
compared with the basic Kerr effect which is generally of order
0.1 \%.
This enhancement occurs because the s polarized light that scatters
from the tip and substrate develops a   
component in the direction perpendicular to the surface that is out of
phase with the s polarized light. This is due
to (1) the
ordinary Kerr effect in which some of the incoming s polarized light 
is converted
to p polarization and (2) the Kerr effect that acts on the
light that is scattered between the tip and the substrate in the near field.
The components of these fields that are perpendicular to the surface and 
lie between the tip and sample are
enhanced by one to two orders of magnitude 
with a corresponding increase in the circular polarization. 

The part of the circular polarization that does not depend 
on the magnetism of the surface is due to light that is reflected from
both the tip and sample surface. The surface reflection
coefficients for s polarized and p polarized light result in phase differences
that produce circular polarized light. 
This circular polarization is 
large and has a characteristic dependence on the observation angle 
which is different from the one 
displayed by the circular polarization that is due to magnetic effects.
This difference in angular dependence may help in separating 
the two contributions.

The resolution of the microscope is roughly 
set by the radius of the tip (probe).
A smaller probe gives a better resolution, however, one must also take into
account that a very small  probe gives a very small scattering cross section.
At the same time, we find that a small probe is essential in obtaining
a high degree of magnetism-dependent polarization of the scattered light. 
Our results indicate that a probe size of about 20 nm is optimal in 
striking the balance between these conflicting requirements.

The paper is organized in the following way.
In section \ref{Theorysec} we present a general theory with which we 
can solve the multiple-scattering problem defined by the model tip-sample
geometry and
calculate the scattered radiation and its polarization state.
This theory uses a spherical model tip. The electromagnetic 
field is calculated taking retardation effects fully into account, 
while the magnetooptic corrections to the reflection coefficients 
of the sample are treated to first 
order in the off-diagonal elements of the dielectric tensor.
In section \ref{Dipolesec}, we discuss the limit of a point dipole 
tip, in order to demonstrate the central physics more clearly, and also to
make a comparison with the case of light emission from an STM probing a
magnetic surface.\cite{APJ}
In Sec.\ \ref{Resultssec} we present and discuss 
the numerical results obtained
using the general theory of Sec.\ \ref{Theorysec}.

\section{Theory}
\label{Theorysec}

\subsection{Basic considerations}
In this section we present a theoretical formalism that allows us 
to calculate the polarization of light scattered from a tip-sample
geometry in which the model tip is a sphere of finite size (the radius of the
sphere corresponds roughly to the radius of curvature of the real tip).
The calculation takes into account higher multipoles on the 
sphere as well as retardation effects.

\begin{figure}
\centerline{
\psfig{file=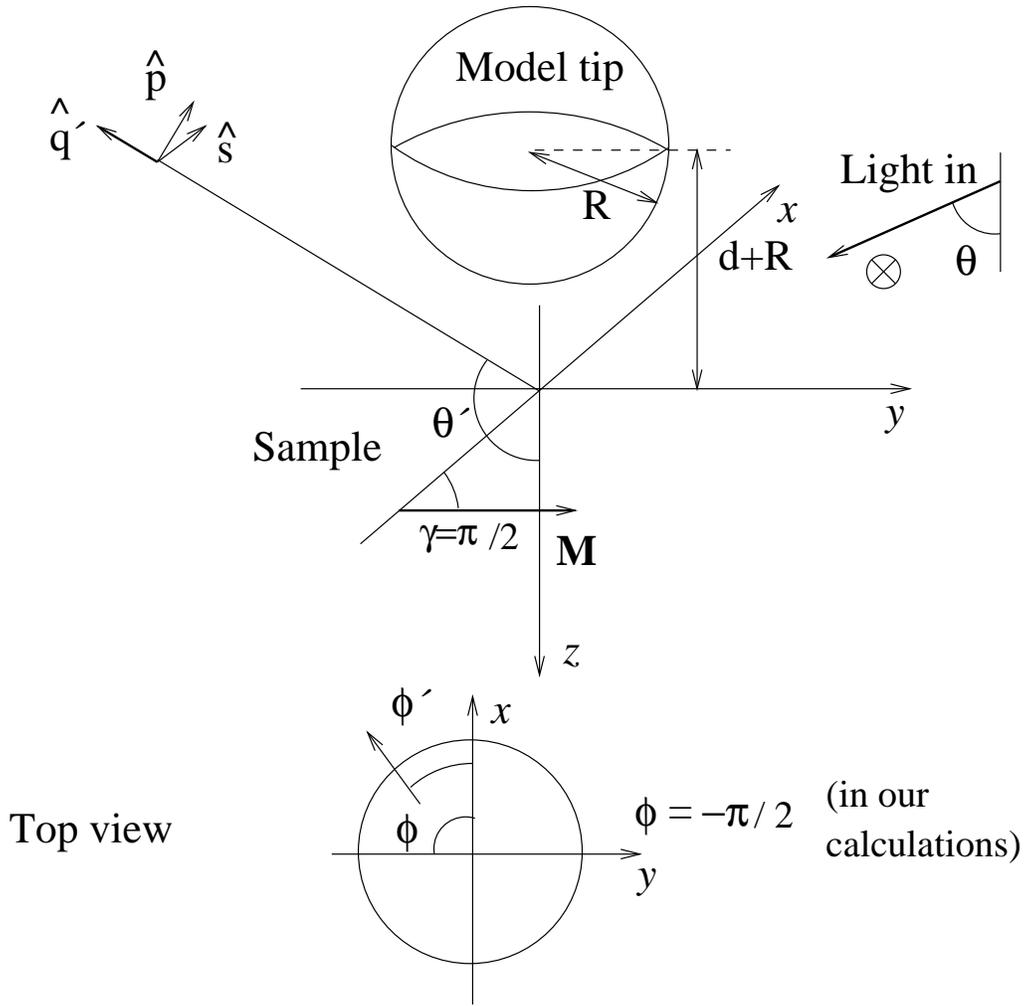,%
bbllx=107 pt,bblly=207 pt,bburx=504 pt,bbury=585 pt}}
\vspace{1 cm}
\caption{
The model geometry used in the calculation.
The vectors $\hat{\bf q}'$, $\hat{\bf p}$, 
and $\hat{\bf s}$ indicate propagation and
polarization directions in the radiation zone. 
$\theta$ and $\phi$ and $\theta'$  and $\phi'$ are the
angles of incidence  and observation, respectively, in a spherical 
coordinate system.
}
\label{geom_fig}
\end{figure}

We consider a situation illustrated in 
Fig.\ \ref{geom_fig} where the magnetic sample fills the half-space
$z>0$, the model tip is a sphere with radius $R$ that has its center
on the $z$ axis at $z=-(d+R)$, thus the tip-sample separation (smallest
distance) is $d$.
A plane wave incident on this system with wave vector  ${\bf q}$ 
can be written as 
(all fields etc.\ vary with time as $e^{-i\omega t}$, but we omit this 
factor here and in the following)
\be
  {\bf E} ({\bf r})= \left\{ E^{(s)} [\hat{\bf z}\crossprod \hat{\bf q}_{\|}]
 - E^{(p)} [\hat{\bf q} \crossprod 
            (\hat{\bf z}\crossprod\hat{\bf q}_{\|})] \right\}
  e^{i{\bf q}\cdot{\bf r}}.
\label{incoming}
\ee
For any given wave vector ${\bf q}$ the polarization vectors for
s and p polarization are, respectively,
$$
 \hat{\bf s}= \hat{\bf z}\crossprod \hat{\bf q}_{\|} 
$$ 
and
$$
 \hat{\bf p}= 
      - \hat{\bf q} \crossprod (\hat{\bf z}\crossprod \hat{\bf q}_{\|}),
$$ 
where ${\bf q}_{\|}$ is the projection of ${\bf q}$ in the surface plane.
Both $\hat{\bf s}$ and $\hat{\bf p}$ are orthogonal to the unit 
vector $\hat{\bf q}$.
We will calculate the radiation sent out in an arbitrary direction
${\bf q}'$ after reflection off the tip-sample system. 
The radiated electric field propagating along ${\bf q}'$ can be written as
\be
  {\bf E}_{\rm rad}= 
  \left\{ E_{\rm rad}^{(s)} [\hat{\bf z}\crossprod \hat{\bf q}'_{\|}]
 - E_{\rm rad}^{(p)} 
  [\hat{\bf q}' \crossprod (\hat{\bf z}\crossprod\hat{\bf q}'_{\|})] \right\}
  e^{i{\bf q}'\cdot{\bf r}}.
\label{radiated}
\ee
The polarization state of this 
radiation is different from that of the incoming light.
First, there is a rather large, in this context less interesting,
contribution to the circular polarization that comes about
because the reflection probability from the surface is different for 
s polarized and p polarized light.
In addition, the scattered light acquires a certain degree of 
circular polarization due to the Kerr effect; of course, for our purposes 
this is the interesting contribution.
An incident $s$ polarized wave will electrically polarize 
the tip in a direction parallel to the sample surface, but
reflections off the magnetic sample surface also gives rise to a field 
component perpendicular to the surface. There is a strong
enhancement of the component of the field perpendicular to the surface
due to the cavity formed between tip and sample, and therefore this second
contribution to the circular polarization should also be detectable. 

The degree of circular polarization of the outgoing radiation is given
in terms of the Stokes parameters as 
\begin{equation} 
 \rho_{\rm CP} = \frac{S_{3}}{S_{0}}. 
\label{stokeseq} 
\end{equation} 
Following Jackson's definitions,\cite{Jackson} the Stokes parameter $S_0$ 
is proportional to the total radiated differential power, whereas 
$S_3$ is proportional to the difference in intensity between left and
right hand circularly polarized light. 
With the geometry illustrated in Fig.\ \ref{geom_fig}, one finds that in
the radiation zone, far away from the tip and the sample surface,
$
 \hat{\bf s}= {\hat{\phi}}'
$ 
and
$
 \hat{\bf p}= \hat{\theta}',
$
where $\hat{\bf \phi}'$ and $\hat{\theta}'$ denote the angular direction of
the scattered light.
 
Left hand circularly polarized light (positive helicity) 
is associated with the unit polarization vector 
$\hat{\epsilon}_{+} = ({\hat{\bf p}}+i{\hat{\bf s}})/ \sqrt{2}$,
while for right hand circularly polarized light (negative helicity) 
$\hat{\epsilon}_{-} = ({\hat{\bf p}}-i{\hat{\bf s}})/ \sqrt{2}$.
For the 
radiation propagating along the ${\bf q}'$-direction 
\be 
 S_0\equiv  
      |\hat{\epsilon}_{+}^*\cdot{\bf E}_{rad}|^2 
    + 
      |\hat{\epsilon}_{-}^*\cdot{\bf E}_{rad}|^2 
   = 
    |E_{\rm rad}^{(p)}|^2 
 + 
    |E_{\rm rad}^{(s)}|^2, 
\ee 
and
\be 
 S_3\equiv  
      |\hat{\epsilon}_{+}^*\cdot{\bf E}_{rad}|^2 
    - 
      |\hat{\epsilon}_{-}^*\cdot{\bf E}_{rad}|^2 
   = 
 -2\, {\rm Im} \left[ 
  E_{\rm rad}^{(s)*} E_{\rm rad}^{(p)} 
 \right].
\ee 
Hence the degree of polarization can be expressed as
\be
 \rho_{\rm CP} = -2\, {\rm Im} 
     \left[
   \frac
{ E_{\rm rad}^{(s)*} E_{\rm rad}^{(p)} }
{ |E_{\rm rad}^{(p)}|^2 + |E_{\rm rad}^{(s)}|^2 } 
       \right] .
\label{rhofraction}
\ee

\subsection{Sample dielectric tensor}

We have expressed the measured magnetic circular dichroism 
in terms of the field amplitudes at the detector. These in turn have
to be calculated with  the tip-sample geometry and tip and sample
dielectric functions  as input.\cite{Nonlocalnote}
Because the sample is magnetic, its dielectric function
is a matrix that takes the form                               
\begin{equation} 
\epsilon_{ij} = \left( \begin{array}{ccc} \epsilon_S(\omega) & 
\epsilon_{1}\cos\varphi & -\epsilon_{1}\sin\gamma \sin\varphi \\ 
-\epsilon_{1}\cos\varphi & \epsilon_S(\omega) & \epsilon_{1}\cos\gamma 
\sin\varphi \\ \epsilon_{1}\sin\gamma \sin\varphi & -\epsilon_{1}\cos\gamma 
\sin\varphi & \epsilon_S(\omega) 
\end{array} \right)  
\label{epstenseq} 
\end{equation} 
where 
$\epsilon_S(\omega)$ is the substrate dielectric function. 
(Sometimes $\epsilon_1$ appearing in the off-diagonal elements
is written $\epsilon_{1}(\omega) \equiv iQ(\omega)\epsilon_S(\omega)$, 
Q being the so called magneto-optical constant.)
The angles $\varphi$ and $\gamma$ relative to the x axis
specify the direction of the magnetization. We will work
in a configuration where $\varphi=\gamma=\pi/2$ which
means that the magnetization is directed 
along the y-axis 
$$
{\bf M} = M{\hat{\bf M}} =M {\hat{\bf y}},
$$
(see Fig.\ \ref{geom_fig}).
Then the dielectric tensor for the sample takes the form \cite{zak}
\begin{equation}
\epsilon_{ij} = \left( 
 \begin{array}{ccc} 
\epsilon_S(\omega) & 0 & -\epsilon_{1}(\omega) \\ 
0 & \epsilon_S(\omega) & 0 \\ 
\epsilon_{1}(\omega) & 0 & 
\epsilon_S(\omega)
\end{array} \right).
\label{epstenssimple}
\end{equation}
In Table \ref{eps1tab} we list values, taken from 
Ref.\ \onlinecite{stearns}, for $\epsilon_1(\omega)$ used in the 
present calculations.

\subsection{Surface response}

When a plane wave of the form in Eq.\ (\ref{incoming}) is reflected 
from the sample surface, ${\bf q}_{\|}$ remains unchanged. The reflected wave
can in general be written as
\be
  {\bf E}_{r}= \left\{ 
   E_{r}^{(s)} {[\hat{\bf z}\crossprod \hat{\bf q}_{\|}]}
 - E_{r}^{(p)} [\hat{\bf q} \crossprod (\hat{\bf z}\crossprod\hat{\bf q}_{\|})] 
 \right\} e^{i{\bf q}_{-}\cdot{\bf r}},
\label{reflected}
\ee
where 
$$
{\bf q}_{-}=({\bf q}_{\|},-p) \ \ {\rm with}\ \  p=\sqrt{k^2-q_{\|}^2}
$$  
and $k=\omega/c$, the subscript $-$
expressing that the reflected wave propagates in the negative $z$ direction.

The reflected amplitude varies linearly with the amplitudes of the 
incident wave.  Thus, for $s$ polarization,
\be
  E_{r}^{s} = \rho_s E^{(s)} + \rho_{sp}'  E^{(p)}.
\label{Erseq}
\ee
The response function $\rho_s$ describes the ``normal''
reflection that takes place whether or not the sample is magnetic.
The Fresnel formula for $s$ polarization is
\be 
 \rho_s = \frac{p-p_S}{p+p_S},
\label{rho_s}
\ee
where  
$$
p_S=\sqrt{k_S^2 -q_{\|}^2}.
$$
$k=\omega/c$, and $k_S=k\sqrt{\epsilon_S(\omega)} $ are the wave-vector
magnitudes in vacuum and the sample material, respectively.
The second term in Eq.\ (\ref{Erseq}) applies to  magnetic surfaces
for which the off-diagonal element $\epsilon_1(\omega)$
in the dielectric tensor
is non-zero and polarization mixing as a result of 
``anomalous'' reflection events becomes possible.
The conversion factor can be explicitly written as
\be
 \rho_{sp}' = \frac{\epsilon_1(\omega)\,\, q_{\|}\, p\, k} 
                               {p_S\, (p+p_S)(\epsilon_S(\omega)\, p +p_S)}
    { [\hat{\bf M}\cdot \hat{\bf q}_{\|}]}
 \equiv
 \rho_L [\hat{\bf M}\cdot \hat{\bf q}_{\|}].
\label{rhol}
\ee
This expression can be derived by the methods developed by 
Zak {\it et al}.\cite{zak}
There is also a contribution from 
the off-diagonal elements in the dielectric tensor to the 
$s$-to-$s$ reflection coefficient. However, this contribution 
is exceedingly small since it depends on the square of
$\epsilon_1(\omega)$ and we neglect it in the following. 

\begin{figure}
\centerline{
\psfig{file=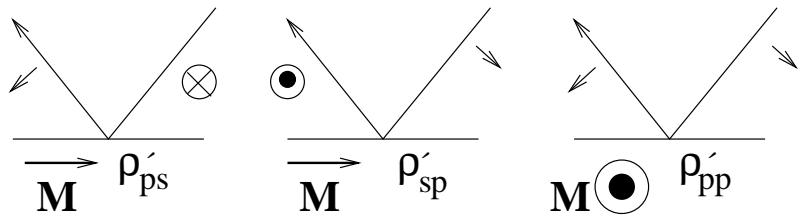,%
bbllx=157 pt,bblly=356 pt,bburx=455 pt,bbury=436 pt}}
\vspace{0.1 cm}
\caption{
Illustration of the different ``anomalous'' scattering processes 
involving the Kerr effect that can take place at the sample surface.
Conversion of $s$ to $p$ polarization and vice versa 
is most effective when the 
in-plane component of the wave vector is collinear with the 
sample magnetization ${\bf M}$. However, the Kerr contribution to
the $p$ to $p$ reflection is most effective when ${\bf q}_{\|}$ and 
${\bf M}$ are perpendicular (NB. In the figure to the right the viewpoint,
but not the magnetization differs from the other figures).
}
\label{process_fig}
\end{figure}

In analogy with Eq.\ (\ref{Erseq}), the reflected $p$ polarized light 
is associated with the electric field
\be
  E_{r}^{p} = \rho_p E^{(p)} + \rho_{ps}' E^{(s)} + \rho_{pp}' E^{(p)}.
\ee
In this case the Fresnel formula is
\be
 \rho_p = \frac{\epsilon_S(\omega)\, p-p_S}{\epsilon_S(\omega)\, p+p_S},
\label{rho_p}
\ee
while the conversion factors originating from $\epsilon_1(\omega)$,
$\rho_{ps}'$ and $\rho_{pp}'$ are given by
\be
 \rho_{ps}' = -\rho_{sp}' =   - \rho_L 
      [{\hat{\bf M}}\cdot {\hat{\bf q}_{\|}}],
\label{rhol2}
\ee
[$\rho_L$ was introduced in Eq.\ (\ref{rhol})] and
\be
 \rho_{pp}' = 
 \frac{2 \epsilon_1(\omega)\, p\, q_{\|}}{(\epsilon_S(\omega)\, p + p_S)^2 } 
  [{\hat{\bf z}}\cdot ({\hat{\bf q}}_{\|}\crossprod {\hat{\bf M}})]
 \equiv
  \rho_T
  [{\hat{\bf z}}\cdot ({\hat{\bf q}}_{\|}\crossprod {\hat{\bf M}})].
\label{rhot}
\ee
Thus, unlike $\rho_{ps}'$ and $\rho_{sp}'$, $\rho_{pp}'$ vanishes when 
${\bf q}_{\|}$ and ${\bf M}$ are parallel.
Figure \ref{process_fig} illustrates the three possible anomalous
scattering processes.

The expressions for all the surface response functions can 
be extended to the case of evanescent
waves for which $q_{\|}=|{\bf q}_{\|}| > k$, and $p$ is imaginary.
(When evaluating the square roots defining $p$ and $p_S$ the branch cut lies
below the positive real axis.)
Later when we deal with the coupling between the sphere and the sample 
this becomes important since the coupling is mediated both by 
propagating and evanescent waves.

\subsection{Scattering from the sphere}
\label{spherescattsec}

We now
deal with the scattering off a sphere whose optical properties 
are described by its dielectric function $\epsilon_T(\omega)$.
To this end we expand the electromagnetic field inside, and just outside
the sphere in terms
of electric (E) and magnetic (M) multipoles.   
We use the same  notation  as Jackson, \cite{Jackson} so that
inside the sphere the electric field  is written
(the spherical coordinate system used here has the origin at the center
of the sphere)
\be 
 {\bf E}=\sum _{lm} k \, c_{lm}^{(M)}\,  j_l(k_T r){\bf X}_{lm}+
 \frac{i}{\epsilon_T}\nabla\crossprod
 \left[c_{lm}^{(E)} j_l(k_T r){\bf X}_{lm}\right],
\ee 
while just outside the sphere the solution is a linear 
combination of incident and 
outgoing waves which can be written
\barr 
 {\bf E}= &&\sum _{lm} 
  k \, [a_{lm}^{(M)}\,j_l(k r) + b_{lm}^{(M)} h_l(kr)]\, {\bf X}_{lm}+
\breakeq
 + i\,\nabla\crossprod
 \left\{[a_{lm}^{(E)} j_l(kr) + b_{lm}^{(E)} h_l(kr)]\, {\bf X}_{lm}\right\}.
\label{Eseries}
\earr 
The corresponding magnetic field outside the sphere is
\barr 
 {\bf B} = &&\sum _{lm} \frac{k}{c} \, 
  [a_{lm}^{(E)}\,  j_l(k r) + b_{lm}^{(E)} h_l(kr)] {\bf X}_{lm}
\breakeq
 -
 \frac{i}{c}\nabla\crossprod
 \left\{ 
  \left[a_{lm}^{(M)} j_l(kr) + b_{lm}^{(M)} h_l(kr) \right]
 {\bf X}_{lm}\right\}.
\label{Bseries}
\earr
Here the vector spherical harmonics
are defined as
\be
 {\bf X}_{lm}(\theta,\phi) \equiv [{\bf L} Y_{lm}(\theta,\phi)]/\sqrt{l(l+1)},
\label{Xlmdef}
\ee
${\bf L} \equiv -i {\bf r} \crossprod \nabla$ 
is the angular momentum operator, and $Y_{lm}$ are 
the usual spherical harmonics.
The vector spherical harmonics form an orthonormal set of functions on the 
surface of a unit sphere.
The functions $j_l$ denote spherical Bessel functions, while  
$h_l\equiv h_l^{(1)}$ 
are spherical Hankel functions  describing outgoing waves.
As before $k=\omega/c$, 
$k_T=\sqrt{\epsilon_T}k$, and $r$
denotes the distance from the center of the sphere.
The coefficients $c_{lm}$, $a_{lm}$, and $b_{lm}$ for the different 
multipoles are as yet unknown; they will be calculated in the following.

Since the angular momentum operator ${\bf L}$ does not have any vector 
component in the radial direction, the same holds true for
${\bf X}_{lm}$. 
Consequently,
for an electric multipole field,
the magnetic field does not have a radial component.
On the other hand the electric field, as one should expect by
visualizing the field from an electric dipole,
has a radial component. 
For a magnetic multipole field, the roles are interchanged;
the magnetic field but not the electric field has a radial component.

In the present context, 
we are primarily interested in knowing how the sphere acts as a scatterer.
It therefore suffices to know what outgoing waves are obtained for given 
incident waves. The ratio between the $b$ and $a$ coefficient for each 
multipole provides this information.
By demanding that the tangential ${\bf E}$ and ${\bf H}$ fields and the 
normal ${\bf B}$ and ${\bf D}$ fields are continuous we arrive at the
following expressions for the sphere ``response'' functions
\be
  s_l^{(E)} \equiv \frac{b_{lm}^{(E)}}{a_{lm}^{(E)}} 
=
 -\, \frac{\epsilon_T\, k R\, j_l'(k R) 
          + j_l (k R)\, (\epsilon_T-1-{\cal J}_l) }
   {\epsilon_T\, k R\, h_l'(k R) 
          + h_l (k R)\, (\epsilon_T -1 -{\cal J}_l) },
\label{slEres}
\ee
and
\be
 s_l^{(M)} \equiv \frac{b_{lm}^{(M)}}{a_{lm}^{(M)}} =
 -
 \frac{ 
k R\, j_l'(k R) 
- 
j_l (k R)\, {\cal J}_l 
}
  { k R\, h_l'(k R) -  h_l (k R)\, {\cal J}_l },
\label{slMres}
\ee
where ${\cal J}_l$ is shorthand for
$$ {\cal J}_l= k_T R\, j_l'(k_TR)/j_l(k_TR). $$
Thanks to the symmetry of the sphere these 
response functions are independent of $m$.

One additional ingredient is needed before we can calculate the field 
around the sphere when it is interacting with the sample. We have to
be able to calculate ``overlap integrals'' between plane waves
and multipole fields and vice versa.
(Note that for lack of a better terminology, 
we will use ``plane wave'' as a common name 
both for propagating plane waves and waves that propagate in the directions
parallel to the sample surface but are evanescent in 
the third, $z$ direction).
A plane wave, like the one in Eq.\ (\ref{incoming}),
impinging on the sphere can  be expanded in terms of regular
multipole contributions [the $j_l$ terms in 
Eqs.\ (\ref{Eseries}) and (\ref{Bseries})].
The resulting $a$ coefficients depend linearly on the amplitudes of the 
incoming wave
\barr
 &&
 a_{lm}^{(E)} 
= 
 f_{lm}^{Ep}({\bf q}) 
 E^{(p)}
+
 f_{lm}^{Es}({\bf q}) 
 E^{(s)}, \ \ {\rm and}
\breakeq
 a_{lm}^{(M)} 
= 
 f_{lm}^{Mp}({\bf q}) 
 E^{(p)}
+
 f_{lm}^{Ms}({\bf q}) 
 E^{(s)}.
\label{afromwave}
\earr
Explicit expressions for the 
proportionality factors  $f$ are derived in the Appendix.

We also have to go in the opposite direction, from a certain multipole
to a plane wave.
In general the field radiated from the sphere can be written as
\barr
 &&  {\bf E}({\bf r}) =
 \sum_{lm,\sigma}
 b_{lm}^{\sigma} 
 \int \frac{d^2 Q_{\|}}{(2\pi)^2}
\, 
 e^{i{\bf Q}\cdot {\bf r}}
\breakeq
\times
 \left[
 g_{lm}^{s\sigma}({\bf Q}) ({\hat{\bf z}}\crossprod {\hat{\bf Q}}_{\|})
 - g_{lm}^{p\sigma}({\bf Q})
  ({\hat{\bf Q}} \crossprod ({\hat{\bf z}}\crossprod{\hat{\bf Q}}_{\|}))
 \right].
\label{planewaveexp}
\earr
Here ${\bf Q} = {\bf Q}_{\|} \pm {\hat{\bf z}}\, \sqrt{k^2 - Q_{\|}^2}$ 
(the sign is chosen according to the direction of propagation or 
exponential decay relative to the $z$ axis)
and $\sigma$ denotes a polarization [(E) or (M)]. 
The factors $g_{lm}^{s\sigma} ({\bf Q})$ and 
$g_{lm}^{p\sigma} ({\bf Q})$ 
are the contributions of a given multipole on the sphere,$(l,m,\sigma)$,
to the amplitude of a plane wave with wave vector ${\bf Q}$.
The explicit calculation of these factors is deferred to the
Appendix.

\subsection{Solution of the multiple-scattering problem}

We are now in a position to solve the multiple-scattering problem
by using the same method as in a 
previous paper.\cite{PJ98}
We solve for the fields at the surface of the
sphere and then calculate the electromagnetic fields elsewhere,
in particular the radiated fields found far from the tip and sample.

In order to carry out  the calculations, it is convenient
to collect the $a$ and $ b$ 
coefficients  of Eqs.\ 
(\ref{Eseries}) and (\ref{Bseries})
into vectors $\vec{a}$ and $\vec{b}$ (in 
``multipole space'') with the structure
\be
  \vec{a} = \left( 
 a_{1-1}^{(E)};
 a_{10}^{(E)}; 
 \ldots;
 a_{l_{max}l_{max}}^{(E)}; 
 a_{1-1}^{(M)};
 \ldots;
 a_{l_{max}l_{max}}^{(M)} 
  \right),
\ee
and similarly for $\vec{b}$.
At the same time a diagonal tensor $\tensor{s}$, can be formed from 
the sphere response functions in Eqs.\ (\ref{slEres}) and (\ref{slMres}), 
so that one can write
\be
 \vec{b} = \tensor{s} \vec{a}.
\ee 

The magnetooptic surface response functions 
$\rho'_{ps}$,
$\rho'_{sp}$, and
$\rho'_{pp}$ are small, and one can therefore treat the magnetooptic effects
by means of a series expansion in $\epsilon_1$.
We begin by calculating the fields to {\em zeroth order in the magnetooptic 
response}.
The field impinging on the sphere is a sum of three contributions
(Figure 3) which
in our vector notation can be written
\be
 \vec{a} = \vec{a}^{\rm dir} + \vec{a}^{\rm ref} + \tensor{N} \vec{a}.
\label{asyst}
\ee
Here
$\vec{a}^{\rm dir}$ represents the amplitude of the field from 
the original incident  wave; 
$\vec{a}^{\rm ref}$  gives the amplitude from the incident wave 
reflected {\em once} from the sample;
and finally, $\tensor{N}\vec{a}$ represent waves sent out from the 
sphere itself that return to it after being reflected from the sample.
Note that since the last term contains the exact field ($\vec{a}$) it accounts
for all multiple scattering events in which light is scattered between 
the sphere and sample an arbitrary number of times.

\begin{figure}
\centerline{
\psfig{file=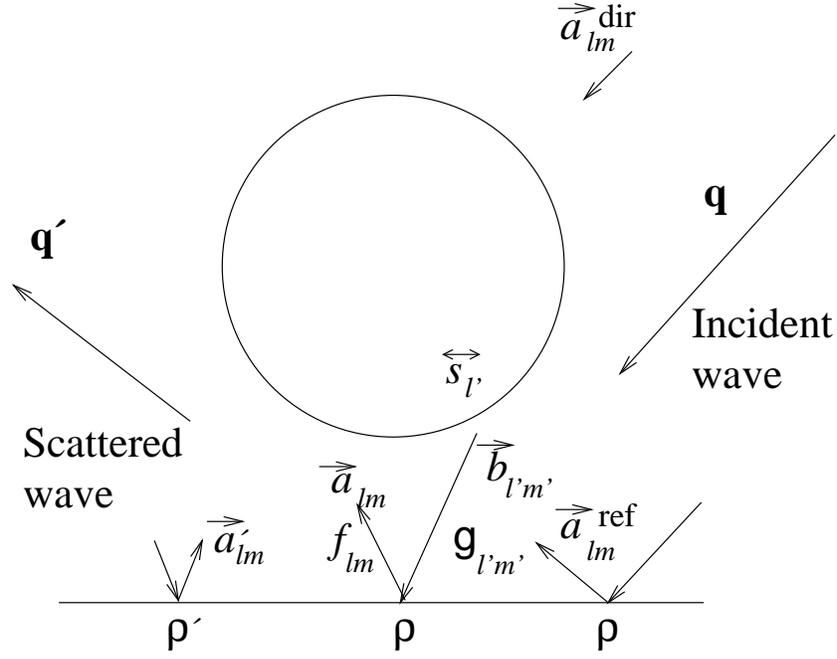,%
bbllx=149 pt,bblly=275 pt,bburx=463 pt,bbury=516 pt}}
\vspace{1 cm}
\caption{
Illustration of the general scheme employed in the calculation.
An incident plane wave can be expressed  as a multipole expansion of
waves impinging on the sphere, either directly ($\vec{a}^{\rm dir}$) or after 
an initial reflection off the sample ($\vec{a}^{\rm ref}$).
Subsequently, multiple scattering takes place 
between the sphere and the sample.
The sphere response functions $s$ determines the strength of 
the outgoing spherical waves $\vec{b}_{l'm'}$, 
which can be expanded into plane waves 
according to the functions $g_{l'm'}$ and after reflection off the sample 
(governed by $\rho$), the functions $f_{lm}$ determine the strengths
$\vec{a}_{lm}$  of the different
multipole contributions to the reflected wave that again impinges 
on the sphere. 
Anomalous (Kerr effect) reflections which are governed by the functions
$\rho'$ generate waves incident on the sphere with strength $\vec{a}'_{lm}$.
}
\label{scheme_fig}
\end{figure}

With an incident wave given by Eq.\ (\ref{incoming}), the elements 
entering $\vec{a}^{\rm dir}$ are found by using Eq.\ (\ref{afromwave})
together with Eqs.\ (\ref{fUeq}) and (\ref{fVeq}).
In an analogous way, the elements of $\vec{a}^{\rm ref}$ are
\barr
 &&
 a_{lm}^{(E),{\rm ref}} 
= 
 f_{lm}^{Ep}({\bf q}_{-}) 
 \rho_p
 E^{(p)}
+
 f_{lm}^{Es}({\bf q}_{-}) 
 \rho_s
 E^{(s)}, \ \ {\rm and}
\breakeq
 a_{lm}^{(M),{\rm ref}} 
= 
 f_{lm}^{Mp}({\bf q}_{-}) 
 \rho_p
 E^{(p)}
+
 f_{lm}^{Ms}({\bf q}_{-}) 
 \rho_s
 E^{(s)}.
\label{arefeq}
\earr
Here $\rho_p E^{(p)}$ and $\rho_s E^{(s)}$  are the amplitudes of the
reflected wave.
The $f$ functions now have ${\bf q}_{-} $ as an argument
since the direction of
propagation  relative to the $z$ axis changes upon reflection.
In the last term of Eq.\ (\ref{asyst}), $\tensor{N}$ is a tensor 
in multipole space.  Its elements describe how 
a wave with angular momentum $l'$ and $m'$ and polarization $\sigma'$
hits the sphere, is reflected, and returns to the sphere again after 
reflection from the plane, now with angular momentum $l$ and $m$ and
polarization $\sigma$.
As a consequence, a tensor element is written
\be
 N_{lm,l'm'}^{\sigma\sigma'} = 
 s_{l'}^{\sigma'}
 \int \frac{d^2Q_{\|}} { (2\pi)^2} \sum_{\sigma''} 
 f_{lm}^{\sigma\sigma''} ({\bf Q}_{-}) \, \rho_{\sigma''}(Q_{\|})
 g_{l'm'}^{\sigma''\sigma'} ({\bf Q}_{+}).
\label{Nintegral}
\ee
In this equation $\sigma''$ stands for either $s$ or $p$ polarization,
and
$$
 {\bf Q}_{\pm} = {\bf Q}_{\|} \pm {\hat{\bf z}} \sqrt{k^2 - |{\bf Q}_{\|}|^2}.
$$
Equation (\ref{Nintegral}) can be derived formally along the lines of the 
calculation presented in Ref.\ \onlinecite{PJ98},
but it can also be understood in a more intuitive way as follows
(see Fig.\ \ref{scheme_fig} for a schematic illustration).
The first factor $s_{l'}^{\sigma'}$, connects the
$a$ coefficient of the initially incident multipole to the corresponding
$b$ coefficient; then the function $g({\bf Q}_{+})$ describes the overlap
between the multipole $l',m',\sigma'$ and a  plane wave, $\rho$ the 
reflection of that plane wave off the sample, and 
$f({\bf Q}_{-})$ gives the overlap between the reflected wave and 
the multipole  $l,m,\sigma$. Because the propagation to the sample surface
and back involves all wave vectors, a  two-dimensional wave-vector
integration is required.
To evaluate Eq.\ (\ref{Nintegral}) it is best to use cylindrical
coordinates in which the angular integration is trivial. Thanks to 
the cylindrical symmetry of the model geometry only tensor elements 
with $m=m'$ are non-zero. 
The remaining $|\bf{Q}_{\|}|$ integration
runs from 0 to $\infty$ so that the sphere-sample
coupling is mediated by both propagating and evanescent waves.

Equation (\ref{asyst}) is solved by a matrix inversion
(of course then only a finite number of multipoles can be retained)
\be 
 \vec{a} = \left[\tensor{1} - \tensor{N} \right]^{-1} \left(
  \vec{a}^{\rm dir} + \vec{a}^{\rm ref} 
  \right).
\label{asolveq}
\ee
Knowing the coefficients $a_{lm}$ we can calculate the electromagnetic 
field everywhere to zeroth order in the magnetooptic response.
The various ingredients entering the calculation of the
scattering processes are shown schematically in
Fig.\ \ref{scheme_fig}.

We next proceed to calculate the corrections to the fields 
to {\em first order in  the magnetooptic response}.
To this end we introduce another set of coefficients $a_{lm}'^{(E)}$
and $a_{lm}'^{(M)}$ that vary linearly with $\epsilon_1$,
the off-diagonal elements of the sample dielectric tensor.
Together $a_{lm}'^{(E)}$ and $a_{lm}'^{(M)}$  form the 
vector $\vec{a}'$ in multipole space.
One can solve for $\vec{a}'$ from the equation
\be 
 \vec{a}' = \vec{a}'^{\rm ext} + \tensor{N} \vec{a}',
\label{aprimsyst}
\ee
in which all terms are first order in $\epsilon_1$.
The first term on the right hand side acts as a source term  for
the first-order-in-$\epsilon_1$ fields,
while the last term 
accounts for  multiple, normal (i.e.\ not involving the Kerr 
effect) reflections between the tip and
sample; the fields appearing there, described by $\vec{a}'$,  have already 
undergone an anomalous reflection.  

There are two contributions to $\vec{a}'^{\rm ext}$,
one due to an anomalous reflection of the incident wave the very first time 
it hits the sample, and one due to waves incident from the sphere that are
anomalously reflected, thus
\be 
  \vec{a}'^{\rm ext} 
=
  \vec{a}'^{\rm ref} 
 + \tensor{K} \vec{a}.
\label{aprimext}
\ee
Both terms on the right hand side are first order in $\epsilon_1$, the first
one being the ordinary Kerr term and the second the tip-induced Kerr term.
In analogy with Eq.\ (\ref{arefeq}),
\barr
 a_{lm}'^{(E),{\rm ref}} 
&&= 
 f_{lm}^{Ep}({\bf q}_{-}) 
 (\rho_{pp}' E^{(p)} + \rho_{ps}' E^{(s)})
\breakeq
+
 f_{lm}^{Es}({\bf q}_{-}) 
 \rho_{sp}' E^{(p)}
\earr
and 
\barr
 a_{lm}'^{(M),{\rm ref}} 
&& = 
 f_{lm}^{Mp}({\bf q}_{-}) 
 (\rho_{pp}' E^{(p)} + \rho_{ps}' E^{(s)})
\breakeq
+
 f_{lm}^{Ms}({\bf q}_{-}) 
 \rho_{sp}' E^{(p)}.
\earr
In these expressions we have suppressed the ${\bf q}_{\|}$ dependence of 
$\rho_{pp}'$, $\rho_{ps}'$, and $\rho_{sp}'$.
This dependence becomes important when evaluating the tensor elements in
$\tensor{K}$,
\barr
 && K_{lm,l'm'}^{\sigma\sigma'} =
 s_{l'}^{\sigma'}
\sum_{\sigma'', \sigma'''} 
\breakeq
\times
 \int \frac{d^2Q_{\|}} { (2\pi)^2} \,
 f_{lm}^{\sigma\sigma''} ({\bf Q}_{-}) \, 
 \rho_{\sigma'', \sigma'''}' ({\bf Q}_{\|}) \,
 g_{l'm'}^{\sigma'''\sigma'} ({\bf Q}_{+}).
\label{Kintegral}
\earr
Here $\sigma''$ and $\sigma'''$ stand for either $s$ or $p$ polarization
(and $\rho_{ss}' =0$ to first order in $\epsilon_1$).
The various factors play the same roles here as in Eq.\ (\ref{Nintegral})
except, of course, for the fact that we now collect
the effects of anomalous reflection events in the sample
described by $\rho_{\sigma''\sigma'''}'$.
As in the previous case the angular integrations can be done 
analytically, and here, due to the angular dependence of 
$\rho_{\sigma,\sigma'}'$, an element in $\tensor{K}$
is nonzero if, and only if, $m=m'\pm 1$.
Once $\tensor{K}$  has been calculated, we evaluate 
$\vec{a}'^{\rm ext}$ and then solve for $\vec{a}'$ 
\be
 \vec{a}' = \left[\tensor{1} - \tensor{N} \right]^{-1} \vec{a}'^{\rm ext}
 = \left[\tensor{1} - \tensor{N} \right]^{-1} 
\left[ \vec{a}'^{\rm ref} + \tensor{K} \vec{a} \right].
\ee

\subsection{Radiated field}

Finally, we calculate the radiated fields, and the degree of polarization.
To this end we add together all contributions to the
radiated field that have scattered from the sphere at one time or another.

To zeroth order in $\epsilon_1$, the field
is found by considering the outgoing waves coming from the 
sphere as well as the corresponding waves reflected (normal
reflection) off the sample surface.   
The specularly
reflected wave that has not interacted with the sphere is ignored since
an experimental measurement that is sensitive to polarization changes
induced by the Kerr effect must avoid this otherwise dominating,
direct contribution.

The radiated field is calculated to first order in $\epsilon_1$   
along the same lines
using the $a'_{lm}$ terms instead of the $a_{lm}$ terms in the 
multipole expansion as a source for the radiation. There is also 
an additional, in practice rather small, 
contribution to the first-order fields that results
from anomalous reflection of zeroth-order waves coming from the 
sphere and scattering off the sample surface a last time before 
going out to infinity.

The field radiated directly from the sphere  
into a direction defined by the angles $\theta'$ and $\phi'$ 
[$(\theta',\phi')\equiv\Omega'$] 
can be calculated by means of a stationary phase approximation yielding
\barr
  {\bf E}_{\rm dir} && =
  -\, \frac{e^{ikr}}{r} 
 \frac{ik\cos{\theta'}}{2\pi} 
  \sum_{lm\sigma}
 s_l^{\sigma} a_{lm}^{{\rm \Sigma},\sigma}
\breakeq
\times
  \left[ 
  \hat{\theta}' 
 g_{lm}^{p\sigma}({\bf q}'_{-}) 
 +
  \hat{\phi}' 
 g_{lm}^{s\sigma}({\bf q}'_{-}) 
  \right].
\label{Edirsphere2}
\earr
Here 
$$ 
 {\bf q}'_{-} = k {\hat{\bf q}}_{-} =
   k\sin{\theta'} ({\hat{\bf x}} \cos{\phi'} + {\hat{\bf y}} \sin{\phi'})
  - k|\cos{\theta'}| {\hat{\bf z}}.
$$
(with an analogous definition, having a positive $z$ component, for 
${\bf q}'_{+}$ used below), and 
$$
 a_{lm}^{{\rm \Sigma},(M)} = a_{lm}^{(M)} + a_{lm}'^{(M)} , \ \ {\rm and} \ \ 
 a_{lm}^{{\rm \Sigma},(E)} = a_{lm}^{(E)} + a_{lm}'^{(E)},
$$
so that the expression contains contributions to both zeroth and first order
in $\epsilon_1$.
The radiation resulting from normal or anomalous reflections from the 
sample surface can be calculated in a way analogous to 
Eq.\ (\ref{Edirsphere2}). 
The only modifications are that one must
(i) multiply by  the appropriate reflection factor, and (ii)
use ${\bf q}'_+$ instead of ${\bf q}'_-$ 
as an argument in
$g$ since the light in this case first propagates towards 
the sample after having left the sphere.

These contributions can now be added together so that the 
polarization $\rho_{\rm CP}$, of the outgoing light  that
has been scattered off the sphere at least once 
can be calculated.
We obtain
\cite{Notesecond}
%
\barr 
 && E_{\rm rad}^{(s)}
 =
  -\,  \frac{e^{ikr}}{r} 
 \frac{ik\cos{\theta'}}{2\pi} 
  \sum_{lm\sigma} 
 s_l^{\sigma} a_{lm}^{{\rm \Sigma},\sigma}
\breakeq
\times \left[
 g_{lm}^{s\sigma}({\bf q}'_{-})
 + \rho_s(k \sin{\theta'})  g_{lm}^{s\sigma}({\bf q}'_{+})
\right.
\breakeq
\left.
 + \sin{\phi'} \rho_{sp}'({\bf q}'_{\|}) g_{lm}^{p\sigma}({\bf q}'_{+})
 \right]
\label{Erads}
\earr
and
%
\barr 
  && E_{\rm rad}^{(p)}
 =
  -\, \frac{e^{ikr}}{r} 
 \frac{ik\cos{\theta'}}{2\pi} 
  \sum_{lm\sigma} 
 s_l^{\sigma} a_{lm}^{{\rm \Sigma},\sigma}
\breakeq
\times \left[
 g_{lm}^{p\sigma}({\bf q}'_{-})
 + \rho_p(k \sin{\theta'})  g_{lm}^{p\sigma}({\bf q}'_{+})
\right.
\breakeq
\left.
 + \cos{\phi'} \rho_{pp}'({\bf q}'_{\|}) g_{lm}^{p\sigma}({\bf q}'_{+})
 + \sin{\phi'} \rho_{ps}'({\bf q}'_{\|}) g_{lm}^{s\sigma}({\bf q}'_{+})
 \right],
\label{Eradp}
\earr
where the terms are organized according to where
the last scattering event takes place.  The first terms describe
waves that come directly from the sphere, the second terms
give the contributions from waves that come from  a final, normal
scattering event from the sample, while the remaining terms originate from
a final anomalous scattering event from the sample.
The degree of circular polarization is then found by inserting these 
expressions into Eq.\ (\ref{rhofraction}).

\section{Dipole limit}
\label{Dipolesec}

For illustrative purposes
it is useful to study the theory developed in the previous section in a
limit where both the sphere radius and the sphere-sample distance are 
much smaller than the wavelength of light.
When these conditions are fulfilled the sphere can be treated as a
point dipole and retardation effects are negligible.

Returning to Eqs.\ (\ref{slEres}) and (\ref{slMres}) we find 
the following limiting
behavior for the sphere response functions when  both
$kR\ll 1$ and $|k_T R| \ll 1$,
\be
  s_l^{(E)} = \frac{i(kR)^{2l+1} (l+1) }{ (2l-1)!! (2l+1)!!}
              \, \frac{\epsilon_T-1}{l \epsilon_T + (l+1)},
\ee
and
\be
  s_l^{(M)} = - \frac{i(kR)^{2l+3} }{ (2l+1)!! (2l+3)!!}.
\ee
Thus the behavior of a small sphere is dominated by its electric-dipole 
response $s_1^{(E)}\sim R^3$.
Therefore, in the rest of this section we will only discuss the electric
dipole excitations of the sphere and the vectors $\vec{a}$ and $\vec{b}$ can 
effectively be reduced to just three components.

We also need  limiting expressions for the coupling matrix elements
$N_{10,10}^{EE}$, $N_{11,11}^{EE}$, and $N_{1-1,1-1}^{EE}$ (the other 
remaining elements of $\tensor{N}$ vanish).
An explicit evaluation of 
$N_{11,11}^{EE}$ in which the large ${\bf q}_{\|}$ limits 
of $\rho_p = (\epsilon_S-1)/(\epsilon_S+1)$, $\rho_s \rightarrow 0$,
as well as $f$ and $g$ are used yields
\be 
  N_{\|}  \equiv
 N_{11,11}^{EE} =  
 N_{1-1,1-1}^{EE} =  
  \frac{\epsilon_T-1}{\epsilon_T+2} \,
          \frac{\epsilon_S-1} {\epsilon_S+1} \, \frac{R^3}{8(d+R)^3}.
\ee
This expression can be obtained from more intuitive reasoning.
If an electric field ${\bf E}=E{\hat{\bf x}}$ polarizes the sphere, a
dipole moment
\be
 {\bf p} = 4\pi\epsilon_0 R^3
 \, \frac{\epsilon_T-1}{\epsilon_T+2} \,  E {\hat{\bf x}}
\ee
is induced
on the sphere, and  there is an accompanying image dipole in the sample
\be
 {\bf p}_{\rm im} = - {\bf p} \, \frac{\epsilon_S-1}{\epsilon_S+1}.
\ee
The electric field created by the image dipole at the center of the
sphere is
\be 
   {\bf E}_{\rm ind} = 
  \frac{\epsilon_T-1}{\epsilon_T+2} \,
   \frac{\epsilon_S-1} {\epsilon_S+1} \, \frac{R^3}{8(d+R)^3} {\bf E}
\equiv N_{\|} {\bf E}.
\ee
This is just the way the matrix elements of $N$ should work;
given an incoming field at the sphere, $N$ generates the fields
reflected from the sphere and scattered back to it by the sample.

If the incident ${\bf E} $ field points 
in the $z$ direction the field from the image dipole,
which now is proportional to $N_{10,10}^{EE}$  is twice as strong
as in the previous, parallel case. Consequently we have,
\be 
 N_{\perp} \equiv
 N_{10,10}^{EE} =  
\frac{\epsilon_T-1}{\epsilon_T+2} \,
          \frac{\epsilon_S-1} {\epsilon_S+1} \, \frac{R^3}{4(d+R)^3}.
\ee

One can also evaluate the $K$ integrals describing the coupling
due to magnetooptic effects. The results can be written
\barr
&& 
 K \equiv
 K_{10,11}^{EE}
 = 
 -K_{11,10}^{EE}
 = 
 -K_{10,1-1}^{EE}
 = 
 K_{1-1,10}^{EE}
=
\breakeq
=
  \sqrt{2}\, \, \frac{\epsilon_T-1}{\epsilon_T +2} \,
    \frac {\epsilon_1}{(\epsilon_S +1)^2} \,
    \frac{R^3}{8(d+R)^3}.
\earr
The other elements of $\tensor{K}$ vanish in this case.

If an s polarized wave with wave vector ${\bf q}$ impinges on the system 
the wave incident on the dipole can be written
\be 
 \vec{a}^{\rm dir} +
 \vec{a}^{\rm ref} =
  - 2\pi  i \sqrt{\frac{3}{4\pi}}   
 \left[ 1+\rho_s({\bf q}_{\|}) \right] 
  \frac{E^{(s)}}{k} 
 \left[
   \begin{array}{c}
     e^{i\phi} \\
      0 \\
     e^{-i\phi} 
   \end{array}
 \right],
\ee
where $\phi$ denotes the azimuthal angle of $\bf q$, the wave vector
of the incident light.
Because $\tensor{N}$ is diagonal it is simple to solve for the full
vector $\vec{a}$ using Eq.\ (\ref{asolveq}),
\be 
 \vec{a} =
  - \frac{2\pi  i}{1-N_{\|}} \sqrt{\frac{3}{4\pi}}   
 \left[ 1+\rho_s({\bf q}_{\|}) \right] 
  \frac{E^{(s)}}{k} 
 \left[
   \begin{array}{c}
     e^{i\phi} \\
      0 \\
     e^{-i\phi} 
   \end{array}
 \right].
\ee
The sphere is only polarized parallel to the surface at this stage 
($a_{10}^{(E)}=0$).

Before proceeding further, let us insert the above result for 
$\vec{a}$ into Eqs.\ (\ref{Erads}) and (\ref{Eradp}) and calculate the
polarization of the light sent out. Since $E_{\rm rad}^{(s)}$ and 
$E_{\rm rad}^{(p)}$ have many factors in common the final result 
can be written
\be
  \rho_{\rm CP} = -2\, {\rm Im} 
 \left[
  \frac{S^* \, P}{|S|^2 + |P|^2}
 \right],
\label{rhonomagn}
\ee
where 
\barr 
  && S= \cos(\phi' -\phi) \left[ 1+\rho_s(q'_{\|}) \right],\ \ {\rm and} 
\breakeq
  P= \sin(\phi' -\phi) \left[ 1-\rho_p(q'_{\|}) \right] \, \cos{\theta'}.
\earr
With appropriate parameter values
Eq.\ (\ref{rhonomagn}) predicts a rather large circular polarization
even if the sample is non-magnetic. If $\rho_{\rm CP}$ given by this
expression is plotted as a function of $\phi'$ one typically obtains
an oscillating curve.  We also note that in this limit,
$\phi'$ and $\phi'+\pi$ yields exactly the same $\rho_{\rm CP}$.
Thus, if we introduce the quantity
\be
\Delta \rho_{\rm CP} (\phi') = \rho_{\rm CP}(\phi') - \rho_{\rm CP}(\phi' +\pi),
\label{asymmdef}
\ee
we find that in the dipole limit, without a Kerr effect,
$\Delta\rho_{\rm CP} = 0$.
This is an important observation since we will see that when the 
sample is magnetic this symmetry is lost.

Due to $\epsilon_1$, there are  two contributions
to $\vec{a}'^{{\rm ext}}$;
the ordinary Kerr effect 
upon reflection of the incident field yields $\vec{a}'^{{\rm ref}}$,
\be 
 \vec{a}'^{{\rm ref}} = 
 -\epsilon_1 \frac{
 q_{\|} p \,
 2\pi \sqrt{3/4\pi} \sin{\phi}
 }
 {p_S (p+p_S) (\epsilon_S p + p_S) }
 \left[
   \begin{array}{c}
     -e^{i\phi} \cos{\theta'} \\
      \sqrt{2} \sin{\theta'} \\
     e^{-i\phi} \cos{\theta'} 
   \end{array}
 \right]
 E^{\rm (s)},
\ee
and the tip Kerr effect yields 
\be
 \tensor{K} \vec{a} = 
 - \frac{4\pi K \sin{\phi}}{1-N_{\|}} \sqrt{\frac{3}{4\pi}} 
  [1+\rho_s(q_{\|})] 
    \frac{E^{(s)}}{k}
 \left[
   \begin{array}{c}
  0 \\
  1 \\
  0
   \end{array}
 \right]
.
\ee
Both contributions are first order in $\epsilon_1$, 
both have a $\sin{\phi}$ dependence on the direction of propagation 
of the incident light, and the Kerr effect induces a dipole moment 
perpendicular to the sample surface.
We also see that resonances in the tip-sample interaction described by
the factor $1/(1-N_{\|})$ can enhance the tip Kerr effect (though 
for a point dipole model for the tip these effects are usually small).

We can now solve for $\vec{a}'$ by making use of
\be 
 \vec{a}' = \left(\tensor{1}-\tensor{N}\right)^{-1} 
\left(
 \vec{a}'^{{\rm ref}} +
 \tensor{K} \vec{a} 
\right),
\ee
where further tip-sample interactions are included.
Next, inserting the sum $\vec{a}^{\rm \Sigma} = \vec{a} + \vec{a}'$ into
Eqs.\ (\ref{Erads}) and (\ref{Eradp}), the polarization is found.

Before analyzing this it is illustrative to see what  result we get for
the magnetic contribution to the circular polarization, 
to lowest order in the ratio of $E_{\rm rad}^{(p)}$ to 
$E_{\rm rad}^{(s)}$, and comparing this to our earlier result in the reverse
situation with light being emitted by tunneling electrons instead.\cite{APJ}
Inserting the dipolar results above for the ordinary 
and the tip Kerr effect into Eqs.\ (\ref{Erads}) and (\ref{Eradp}) 
we obtain to lowest order, retaining only magnetic contributions
to the circular polarization
\barr
  && \rho_{\rm CP}^{M}
 \approx
     {\rm Im} 
     \left[
    \frac{ 1+G_{\perp} } { 1+G_{\|} }
    \left(
   \frac{-\rho_L(\omega)}{1+\rho_s (q_{\|}) } - 
   \frac{ D(\omega)G_{\parallel} } { \sin{\theta'} }
       \right)
    \frac{ 1+\rho_p (q_{\|}') }{ 1+\rho_s (q_{\|}') }
     \right]
\breakeq
\times
   2\, 
 \frac{ \sin{\phi}\sin^2{\theta'} } { \cos{(\phi-\phi')} },
\label{rhodipmcd}
\earr
where $\rho_L(\omega)$ is defined in Eq.(\ref{rhol})
and\cite{Dnote} 
\be
 D(\omega) = \frac{2\epsilon_1(\omega)}{\epsilon_S^{2}(\omega)-1}.
\ee
Furthermore the image factors G are defined through
\be
 G_{\|,\perp} =
 N_{\|,\perp} 
  /
 (1 - N_{\|,\perp}) 
\label{Gdef}
\ee
for the parallel and perpendicular cases, respectively.
The result for the circular polarization 
of light in the corresponding spontaneous emission 
STM configuration is
\be
  \rho_{\rm CP}^{\rm STM} \approx 2 \, {\rm Im} 
     \left[
       \left(
   \frac{-\rho_L(\omega)}{1+\rho_s} + 
    \frac{D(\omega) G_{\parallel} }{\sin{\theta'}}
       \right)
   \frac{1+\rho_s}{1+\rho_p}
       \right].
\ee
The structure is very similar to that of $(3.17)$ but 
there are a couple of important differences.
For the light-in situation the whole
expression is basically larger by the factor $G_{\perp}$. 
This is one reason  
for why $\rho_{\rm CP}^{M}$ can be
several orders of magnitude larger than $\rho_{\rm CP}^{\rm STM}$; 
the perpendicular 
field enhancement, $G_{\perp}$, can
be very large in the junction between tip and sample, while 
$G_{\parallel}$ is of order unity.  
Moreover, in the last factor $(1+\rho_s)$ and $(1+\rho_p)$ have changed 
places.  This is also an important reason for the large values 
reached by $\rho_{\rm CP}^{M}$.  At a metal surface $(1+\rho_s)$ is usually 
considerably smaller than 1.  

Continuing the analysis of the full results we find that they can be 
summarized as
\be
  \vec{a} =  
 \left[
   \begin{array}{c}
   (- i a_x  + a_y)/ \sqrt{2} \\
  a_z \\
  (i a_x + a_y)/ \sqrt{2}
   \end{array}
 \right],
\ee  
for  $\vec{a}$,
where $a_x$, $a_y$ and $a_z$ are proportional to the induced dipole moments 
in the $x$, $y$, and $z$ directions, respectively.
Let us assume from now on that the incident wave propagates in the 
negative $y$ direction, so that $\phi = -\pi/2$.
This means that $a_x$ will be large, whereas $a_y$ and $a_z$ are much smaller
since they originate from the Kerr effect.
Therefore looking at the radiation sent out in
two opposite directions ($\phi'$ and $\phi'+\pi$) we see that the 
contributions to
$E_{\rm rad}^{(p)}$ and  
$E_{\rm rad}^{(s)}$
caused by the $x$ and $y$ components of the dipole moment 
both change sign upon 
changing the angle of observation, leaving the 
polarization state unchanged as we have already discussed.  
However, the new feature, the dipole moment in
the $z$ direction adds the same contribution to 
$E_{\rm rad}^{(p)}$ in both directions, thereby leading to an asymmetry
in the angular dependence of the polarization.   
Anomalous reflection of the radiation sent out as a result of the
dipole moment oscillating in the $x$ direction also yields contributions
to $E_{\rm rad}^{(s)}$ and $E_{\rm rad}^{(p)}$ 
that do not change sign when changing $\phi'$ to $\phi'+\pi$.
From this follows that the magnetism-related contribution to the 
circular polarization has a characteristic angular dependence
which, as we will see, persists also when going beyond the dipole model.

\section{Results and discussion}
\label{Resultssec}

\subsection{Numerical results}

In this section we will present results of our numerical calculations.
We have solved the multiple-scattering problem at hand using the 
theory developed in Sec.\ \ref{Theorysec}.  In these calculations 
we retained multipoles for which $l \leq 30$ and $|m| \leq 3$.
This means that we get results that are essentially numerically exact;
the relative accuracy obtained for 
$\Delta\rho_{\rm CP}$ is better than 
$10^{-5}$ when $R=20$ nm and $d=0.5$ nm.
If $l_{max}$ is reduced to 10 the relative errors are about 1 \%, and with
$l_{max}=5$ they amount to 5--10 \%.

\begin{figure}
\psfig{figure=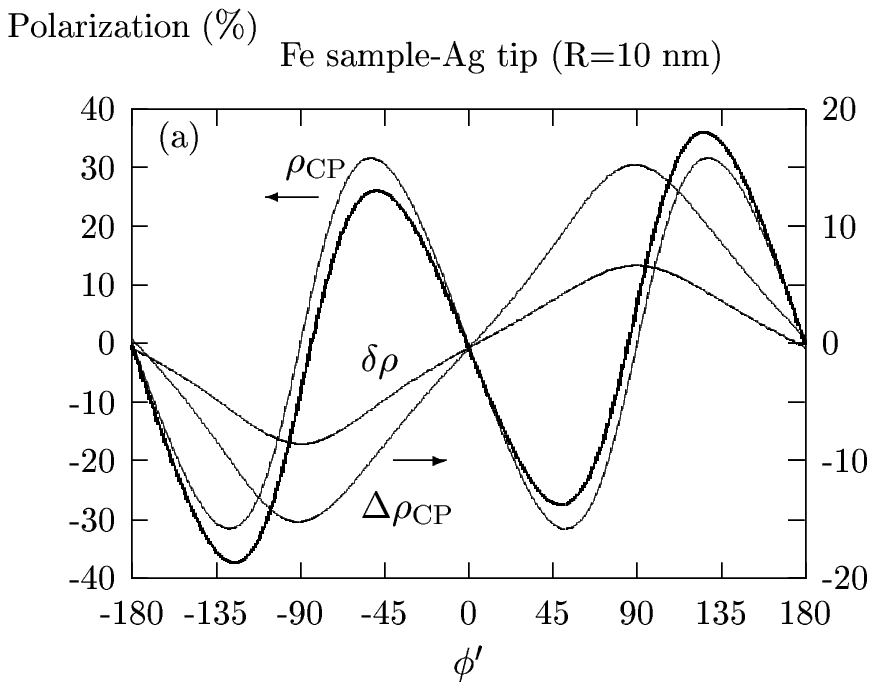}
\vspace{1 cm}
\psfig{figure=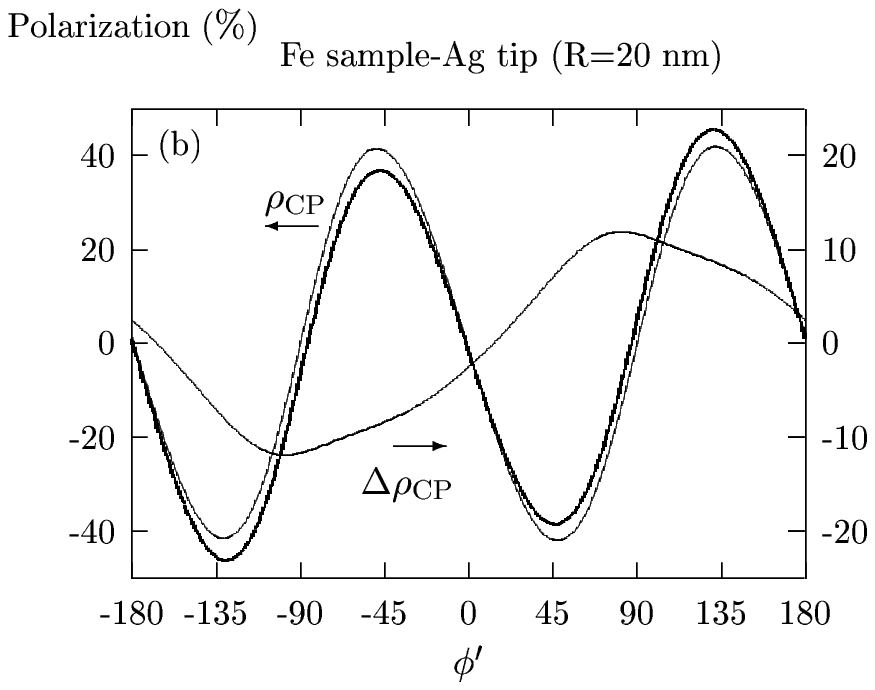}
\vspace{1 cm}
\psfig{figure=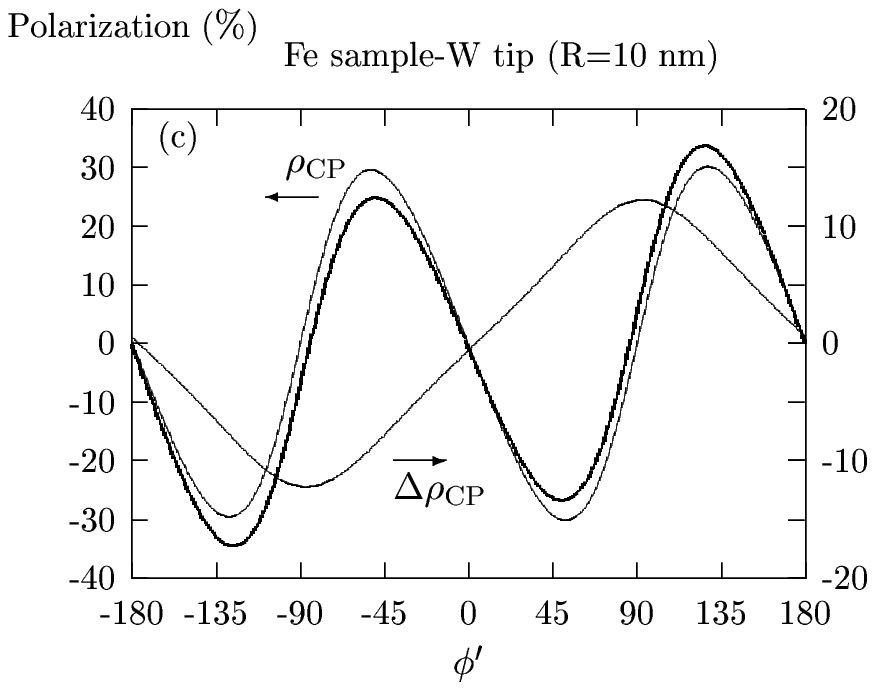}
\vspace{1 cm}
\caption{
Calculated degree of polarization for a Ag tip probing a Fe sample.
The curves in panel (a) were calculated with a tip radius 
$R=$ 10 nm, the ones in panel (b) with $R=$ 20 nm. 
For comparison results calculated with a W tip with 
$R=$ 10 nm are displayed in panel (c). In all cases the tip-sample separation
$d=$0.5 nm.
The photon energy was set to 1.6 eV, $\phi=-90^{\circ}$, 
and the polar angles of incidence and  observation $\theta$ and $\theta'$,
respectively,  both  lie 1 radian off the surface normal
(i.e.\ $\theta$ = 1 rad.\ while $\pi-\theta'$ = 1 rad.).
The two curves with the highest amplitude show the results for 
$\rho_{\rm CP}$.
The thinner of these curves was calculated without accounting
for the magnetooptic
properties of the sample, whereas they have been included in the calculation
yielding the thicker curve.
The remaining curves display (i)
the quantity $\Delta\rho_{\rm CP}$ defined in 
Eq.\ (\protect\ref{asymmdef}), and (ii), in panel (a),
the difference $\delta\rho$
between the results with and without magnetooptic effects.
}
\label{fdeffig}
\end{figure}

Figure \ref{fdeffig} shows results for the 
polarization $\rho_{\rm CP}$ as a function of the observation angle 
$\phi'$.
These results were obtained from a calculation  using a silver tip and an
Fe sample.
For comparison, we have also performed a calculation using a 
W tip [panel (c)]. 
The photon energy was 1.6 eV and both the incident and the scattered 
light propagates in directions forming an angle
of 1 radian with the surface normal.
This means that $\theta=$1, while $\theta'=\pi$-1.

Let us begin the discussion by looking at the results 
in Fig.\ \ref{fdeffig} (a), where the Ag tip radius was set to 10 nm.
Results are shown both with (thick curve) and without (thin curve) 
the magnetooptic effects;  
the two curves are very similar indicating that ``geometric effects''
cause the dominating contribution to the 
polarization $\rho_{\rm CP}$.
However, a closer comparison reveals that the results
including magnetooptic effects show a less symmetric angular
dependence than those obtained without taking magnetooptic effects 
into account.
Therefore magnetooptic effects can be isolated more effectively if the 
quantity $\Delta\rho_{\rm CP}$ defined in Eq.\ (\ref{asymmdef}) is 
plotted.
We see that $\Delta \rho_{\rm CP}$ has peak values exceeding 10 \%,
and at the same time the polarization shows a simple, characteristic 
variation with the observation angle $\phi'$.
It appears that these effects are large enough to be detectable.
We have also plotted the quantity $\delta\rho$, i.e.\ the change
in $\rho_{CP}$ due to magnetooptic effects.  The two curves
are very similar in shape.

In Fig.\ \ref{fdeffig} (b) the tip radius was increased to 20 nm.
The tip is now large enough
that phase differences between light scattered off different parts
of the tip influence the polarization.
This gives $\Delta \rho_{\rm CP}$ a more complicated angular
behavior, and it would be somewhat more difficult to pick out the part of
the signal originating from the Kerr effect.
For  an even larger tip radius this effect  
becomes more pronounced.

There are two different magnetism-related processes that
contribute to the circular polarization of the scattered light.
We already mentioned this in connection with Eqs.\ (\ref{aprimext}) and
(\ref{rhodipmcd}), and the same distinction was also made in 
our previous work on light emission.\cite{APJ}
To begin with there is an ordinary Kerr effect in which
the polarization conversion,
governed by the coefficient $\rho_{ps}'$, takes place primarily when 
the incident wave first hits the sample  surface.  
This induced p polarized wave 
is subsequently enhanced in the cavity between tip and sample;
both tip and sample screen the electric field, surface charges are induced on
both of them and this generates secondary fields, etc. 
In the other process, which we have called the tip Kerr effect, 
the incident wave is reflected back and forth
between tip and sample a number of times and an electric field along the $z$
axis is built up because one of the scattering events off the sample 
is anomalous.  In this case the anomalously reflected wave typically is 
evanescent since it describes the near field around the tip.
The conversion factor $\rho_{pp}'$ dominates these processes,
because, as can be seen from Eqs.\ (\ref{rhol}), (\ref{rhol2}), 
and (\ref{rhot}), unlike  $\rho_{ps}'$ and $\rho_{sp}'$ 
it has a non-zero limit when $q_{\|}\gg k$. 

In the full (multipole) calculations the relative importance of the 
two processes can be estimated, for example, by temporarily eliminating
one or the other contribution.  We find that the ordinary Kerr 
effect in the present case is responsible for some two thirds
of the magnetic contributions to the circular polarization of the 
scattered light. 
In the calculation addressing emission of circularly
polarized light\cite{APJ} we found the opposite situation;
the tip Kerr effect dominated then.
It is relatively straightforward to understand this difference qualitatively:
Near-field effects are considerably more important in the case 
of light emission
because in that experiment one measures how much radiation is produced
by a source, the tunnel current, that is basically localized to a 
nanometersized region in space.
In the present calculation we instead probe the degree to which 
light is scattered 
off the tip-sample system; the near-field effects play a role also here
but it is not as pronounced.

We have concentrated on studying the polarization properties 
of the scattered light at relatively low photon energies ($< 2$ eV)
because $\epsilon_1$ for the magnetic materials 
takes the largest values there. 
In that case the results obtained with a W tip (shown in panel c) and with
a Ag tip (panel a) are not too different from each other.
This result may at first seem surprising given that Ag is a
much better conductor than W and has well-defined surface plasmon excitations.
However, for the photon energies dealt with here 
the field enhancement does not occur as a result of 
truly {\em resonant} interactions with interface plasmons.
Instead the electric field in the tip-sample cavity is enhanced by
a less dramatic mechanism.  
Both Ag and W screen the electric field and therefore surface charges 
that interact with each other are built up on the tip and sample. 
At higher photon energies a Ag tip certainly
gives a stronger field enhancement  than a W tip,
but then again, since $\epsilon_1$ for the magnetic materials 
is much lower there, 
this frequency range is of limited interest in the present context.

\begin{figure}
\psfig{figure=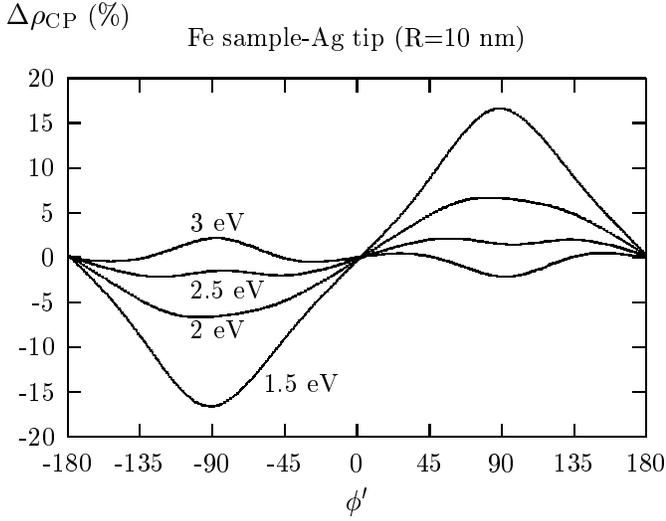}
\vspace{1 cm}
\caption{
The polarization asymmetry $\Delta\rho_{\rm CP}$, defined
in Eq.\ (\protect\ref{asymmdef}),  
plotted as a function of 
 observation  angle $\phi'$,
for a number of photon energies
for a Fe sample scanned by a Ag tip 
with a radius of 10 nm, and with $d$ set to 0.5 nm.
}
\label{plattfig}
\end{figure}

Figure  \ref{plattfig} illustrates in more detail
how the polarization properties of the 
scattered light varies with photon energy.
The highest degree of polarization is reached at 1.5 eV\@.
For photon energies beyond 2 eV the Kerr effect has a rather
small influence on the polarization, the main reason for this is 
that the magnitude of $\epsilon_1$ for Fe decreases with increasing
$\hbar \omega$.

\begin{figure}
\psfig{figure=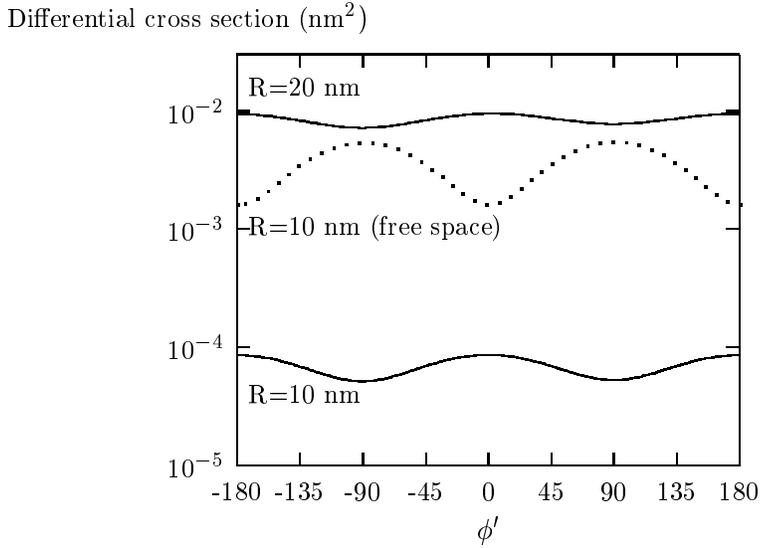}
\vspace{1 cm}
\caption{
Calculated differential scattering cross section for a Ag sphere in front of 
a Fe sample (full curves), and in free space (dotted curve)
as a function of azimuthal observation angle.
The photon energy is 1.6 eV, $d=$ 0.5 nm,
the sphere radius is 10 nm (lower full curve
and dotted curve), and 20 nm  (upper full curve), respectively.
The polar angles of incidence and  observation $\theta$ and $\theta'$,
respectively,  both  lie 1 radian off the surface normal
(i.e.\ $\theta$ = 1 rad.\ while $\pi-\theta'$ = 1 rad.).
}
\label{crossecfig}
\end{figure}

In Fig.\ \ref{crossecfig}
we plot the differential scattering cross section for a Ag tip 
with a radius of either 10 nm or 20 nm. 
As a reference we have also plotted the result
obtained for a Ag sphere with radius 10 nm  in free space.
The sphere in front of an Fe sample has a much smaller cross section
than the one in free space because of destructive interference between 
the waves sent out directly from the sphere and those reflected from the 
sample, which in this case can be thought of as originating primarily from an
image sphere. The radiation patterns differ also between the two
cases. An isolated sphere displays a dipole pattern, but with the sample
present, the angular distribution of radiation mainly follows a quadrupole 
pattern.
To see what the cross sections mean in terms of photon counts, 
let us assume that a 
laser with an intensity of 10$^3$ W/cm$^2$ is illuminating the tip. 
This is equivalent to a photon flow of $\sim 10^8$ photons/(sec.\ nm$^2$).
In the case of the 10-nm-sphere in front of the Fe sample, a detector 
covering a solid angle of 0.1 steradian would collect $\sim 10^3$ 
photons per second.  
This intensity should be enough to perform a polarization
analysis.

\subsection{Experimental implications}

The measurement of the magnetic contribution to the circular
polarization should be feasible in the sense
that the scattered light intensity is sufficient 
and the degree of circular polarization of the scattered light is
large enough to be detectable with existing optical 
techniques. Furthermore, it should at least be possible to reach 
a resolution of 10 nm. However, the removal of the background circular
polarization which is not magnetic in origin is a nontrivial
experimental problem.  The Kerr contribution varies with
scattering angle in a way which is different from other 
nonmagnetic contribution to the polarization,
provided the tip is not too large. This should help in separating
the two contributions to the circular polarization.
It may also be possible to use the fact that the Kerr contribution
to the circular polarization is more sensitive to the tip-sample 
distance than is the background circular polarization, for example by
letting the tip vibrate back and forth. To be specific, 
increasing the tip-sample distance from 0.5 nm to 5 nm 
in Fig.\ \ref{fdeffig} (a) decreases the Kerr contribution 
($\Delta\rho_{\rm CP}$) by about a factor of 2, whereas 
the background contribution is essentially unchanged.
Varying the photon energy could be 
yet another way of differentiating between Kerr and background 
contributions.

In view of the large background, the size of the tip would be a major 
concern in designing an experiment in practice. 
If a traditional STM tip made from a metallic wire is used, regardless
of how sharp the tip is, scattering will take place everywhere 
in the tip shaft.  This scattering will typically increase the 
geometric-background contribution to the circular polarization,
and make it more difficult to detect the magnetic properties of 
the surface. Of course, the incident radiation would be focused to a 
relatively small part of the tip, but this may not be enough to obtain a 
high contrast between the contributions to the polarization 
induced by the Kerr effect and geometric effects, respectively.

As we have indicated in the Introduction,
one possible way of bypassing this difficulty could be  to follow  a
scheme closely related to that introduced by Silva and 
Schultz,\cite{silva} using a small, isolated  metallic particle 
on the surface of the magnetic sample as the probe.  
In order to scan
different parts of the surface, the metallic particle would have to be moved
around using an AFM.  This kind of nanoparticle manipulation has already been
achieved in other contexts.\cite{Junno}

Pufall, Berger, and Schultz\cite{Pufall} have measured 
the Kerr rotation of light scattered from Ag particles placed on a
magnetic substrate.  They found a Kerr rotation that had qualitative
features (in terms of observation-angle dependence) in common with the
results presented here.  The measured Kerr rotation, however, was less than 
1 mrad (corresponding to  a magnetism-related degree of polarization
of 0.1 \% or less).   
It appears from Ref.\ \onlinecite{Pufall} that particles of a 
radius of at least 50 nm were used in the experiment and
at the same time the incident laser light was tuned to the scattering
resonance of the particles at a wavelength of  460 nm ($\approx$ 2.7 eV).
While both of these choices ensure that the scattering cross section 
is much larger
than with the parameter values used in the calculations above, this is
achieved at the cost of a very small degree of  magnetism-related
polarization of the scattered light.
First of all $\epsilon_1(\omega)$ in common magnetic metals
such as Fe and Co decreases with increasing photon energy in
the range of visible light (cf.\ Fig.\ \ref{plattfig}).
Secondly, with a larger particle the in-plane dipole becomes more 
effective in sending out radiation because the image sphere is further
away from the real sphere. As a result the scattering cross section but also 
the  nonmagnetic contributions to $\rho_{\rm CP}$ increase dramatically.
Indeed, using a silver particle of radius $R=50$ nm on a Fe substrate
and a photon energy of 3 eV in our calculations we find that
the magnetism-related contribution to the light polarization 
(i.e.\ $\delta\rho$) is only $\approx $ 0.1 \%. With $R=$50 nm and
a photon energy of 1.6 eV, $\delta\rho\approx$ 2 \% at most, to be compared 
with 6--7 \% for $R=10$ nm (see Fig.\  \ref{fdeffig} (a)).
Thus, based on the results found here we recommend that 
experiments of this kind should use smaller photon energies 
(1.5 eV or so) and smaller particles ($\sim$ 20 nm) than before.

\acknowledgments
We would like to thank M. L. Cohen, R. J. Celotta, L. Montelius, 
and L. Samuelsson for useful discussions.
SPA acknowledges a grant from Iberdrola S.A. 
The work of PJ and SPA
is supported by the Swedish Natural Science Research Council, and by
the European Union through TMR contract ERB--FMRX--0198.

\appendix


\section*{}

In this Appendix, we derive expressions for the functions $f$ and 
$g$ that were introduced in Sec.\ \ref{spherescattsec}.
With a plane wave impinging on the sphere, the full electric field 
outside the sphere can be expanded as in Eq.\ (\ref{Eseries}),
and the regular ($j_l$) terms describe the incoming wave.
We will calculate the
proportionality factors  $f$  appearing in Eq.\ (\ref{afromwave}),
and therefore we first need to evaluate the $a$ coefficients in
Eq.\ (\ref{Eseries}).
It is clear from  Eq.\ (\ref{Eseries}), in view 
of the orthogonality relations 
\be 
 \int d\Omega\, {\bf X}_{l'm'}^*(\Omega)\cdot{\bf X}_{lm}(\Omega)
   =\delta_{ll'}\delta_{mm'}
\label{orthoXX}
\ee
and
\be 
 \int d\Omega\, {\bf X}_{l'm'}^*(\Omega)\cdot
  \nabla \crossprod [(a j_l(kr) +b h_l(kr)) {\bf X}_{lm}(\Omega)]
   =0,
\label{ortho0}
\ee
for the 
vector spherical harmonics, that
\be
 k a_{lm}^{(M)} j_l(kr) = \int  d \Omega\, \, ({\bf X}_{lm})^* \cdot {\bf E}.
\ee
To calculate $a_{lm}^{(E)}$, it is easier to use the overlap between 
the ${\bf B} $ field and a vector spherical harmonic,
\be
 \frac{k}{c} a_{lm}^{(E)} j_l(kr) = 
  \int  d \Omega\, \, ({\bf X}_{lm})^* \cdot {\bf B}.
\ee
To evaluate these surface integrals, 
we use the expansion of a scalar plane wave
(i.e.\ the factor $e^{i{\bf q}\cdot{\bf r}}$) in terms of spherical harmonics
and insert this in Eq.\ (\ref{incoming}) yielding
\barr
 && {\bf E} ({\bf r}) = 
        \left\{ E^{(s)} [{\hat{\bf z}}\crossprod \hat{\bf q}_{\|}]
 - E^{(p)} [{\hat{\bf q}} \crossprod 
    (\hat{\bf z}\crossprod\hat{\bf q}_{\|})] \right\}
 e^{i{\bf q}\cdot{\bf R}_{\rm sph}}
\breakeq
\times
   4\pi\, \sum_{l=0}^{\infty}
   i^l j_l(kR)\sum_m (-1)^m Y_{l,-m}(\Omega)
   Y_{l,m}(\Omega_q).
\label{scalarexp}
\earr
The incident $B$ field can be expanded in a similar way.
The factor $e^{i{\bf q}\cdot{\bf R}_{\rm sph}}$ compensates for the fact
that we here use a coordinate system with the origin at the center
of the sphere [${\bf R}_{\rm sph}= -\hat{\bf z}\,(R+d)$].
The composite variables $\Omega$ and $\Omega_q$ denote the  directions
of $\hat{\bf r}$ (i.e.\ the angles $\theta$ and $\phi$) and
$\hat{\bf q}$ ($\theta_q$ and $\phi_q$).
Equation (\ref{scalarexp}) is also valid for evanescent waves that decay
exponentially in the positive or negative $z$ direction
in which case $\cos{\theta_q}$ is purely imaginary.  From 
the definition in Eq.\ (\ref{Xlmdef}) and the fact that ${\bf L}$ 
can be written in terms of ladder operators as
$$
 {\bf L} 
 =
 \frac{1}{2} (\hat{\bf x}-i\hat{\bf y}) L_{+}  +
 \frac{1}{2} (\hat{\bf x}+i\hat{\bf y}) L_{-}  +
 \hat{\bf z}L_z,
$$
it is clear that ${\bf X}_{lm} $ 
can be expressed in terms of the spherical harmonics
$Y_{l,m+1}$, $Y_{l,m}$, and $Y_{l,m-1}$. 
Combining this with the expansion in Eq.\ (\ref{scalarexp}) yields the
overlap integrals
\be 
 f_{lm}^{Ep}({\bf q})= f_{lm}^{Ms}({\bf q}) = 
  k^{-1} \, U_{lm} ({\bf q}) \,
   e^{i{\bf q}\cdot{\bf R}_{\rm sph}},
\label{fUeq}
\ee
and
\be 
 f_{lm}^{Es}({\bf q}) = \, -\, 
 f_{lm}^{Mp}({\bf q}) 
=
  k^{-1} \, V_{lm} ({\bf q})\, 
   e^{i{\bf q}\cdot{\bf R}_{\rm sph}}.
\label{fVeq}
\ee
Here we have introduced
\barr
  U_{lm}({\bf q}) &&= 
  - \frac{2\pi i^l (-1)^m}  {\sqrt{l(l+1)}} 
   \left[
 \xi_{+}\, F_{+}(l,m)\, Y_{l,-m-1}(\Omega_q)\,
 + \right.
\breakeq
 \left.   +
 \xi_{-}\, F_{-}(l,m)\, Y_{l,-m+1}(\Omega_q)\,
 \right],
\label{Ulmeq}
\earr
where the polarization vector 
\barr
  && \hat{\xi} = \hat{\bf z}\crossprod\hat{\bf q}_{\|} =
  \hat{\bf y} \cos{\phi_q} - \hat{\bf x} \sin{\phi_q},  
  \ \  {\rm with}
\breakeq
 \xi_{\pm} \equiv \xi_x \pm i \xi_y = \pm i e^{\pm i \phi_q}.
\earr
In a similar way
\barr
 && V_{lm}({\bf q}) = 
  - \frac{2\pi i^l (-1)^m}  {\sqrt{l(l+1)}} 
   \left[
 \eta_{+}\, F_{+}(l,m)\, Y_{l,-m-1}(\Omega_q)\,
 + \right.
\breakeq
 \left.   +
 \eta_{-}\, F_{-}(l,m)\, Y_{l,-m+1}(\Omega_q)\,
 - \eta_z\, 2m\, Y_{l,-m}(\Omega_q)
 \right],
\label{Vlmeq}
\earr
where the polarization vector
\barr
 && \hat{\eta}=
        k^{-1} {\bf q} \crossprod (\hat{\bf z}\crossprod\hat{\bf q}_{\|}) 
\breakeq
 \ \  = \hat{\bf x}(-\cos{\theta_{q}} \cos{\phi_q})
 + \hat{\bf y}(-\cos{\theta_{q}} \sin{\phi_q})
 + \hat{\bf z}\sin{\theta_{q}}, \ \ {\rm with}
\breakeq
 \eta_{\pm} \equiv \eta_x \pm i \eta_y = -\cos{\theta_q} e^{\pm i \phi_q}.
\earr
In both Eq.\ (\ref{Ulmeq}) and Eq.\ (\ref{Vlmeq}), 
$F_+ $ and $F_-$ are shorthand symbols  for the 
numerical factors produced by the ladder operators $L_+$ and $L_-$, thus
\barr
&&
 F_{+}(l,m)= \sqrt{(l-m)(l+m+1)},  \ \ {\rm and}
\breakeq
 F_{-}(l,m)= \sqrt{(l+m)(l-m+1)}.
\label{Fdef}
\earr
Finally, we note that the $\phi_q$ dependence of both $U_{lm}$ and 
$V_{lm}$ follows
\be
 U_{lm}({\bf q}) \propto e^{- i m \phi_q},\ \ 
 V_{lm}({\bf q}) \propto e^{- i m \phi_q}.
\label{UVang}
\ee 
The same is true for $f_{lm}$ 
(provided ${\bf R}_{\rm sph}$ lies on the $z$ axis).

Next we must deal with the overlap going in the other direction,
from a given multipole to a plane wave.  
The vector equivalents of the Kirchoff integrals
provides one way of doing this. They read
\barr
 {\bf E}({\bf r})=&&
 {\bf E}^{\rm ext}({\bf r}) +
  \int dS'
  \left[
         ikc(\hat{\bf n}'\crossprod{\bf B}({\bf r}'))G({\bf r},{\bf r}')
\right.
\breakeq 
\left.
         + (\hat{\bf n}'\crossprod{\bf E}({\bf r}'))\crossprod\nabla'G
         + (\hat{\bf n}'\cdot{\bf E}({\bf r}'))\nabla'G
                                       \right]
\label{KirchE}
\earr
and 
\barr
 {\bf B}({\bf r})=&&
 {\bf B}^{\rm ext}({\bf r}) +
  \int dS'
  \left[
 - \frac{ik}{c} (\hat{\bf n}'\crossprod{\bf E}({\bf r}'))G({\bf r},{\bf r}')
\right.
\breakeq 
\left.
         + (\hat{\bf n}'\crossprod{\bf B}({\bf r}'))\crossprod\nabla'G
         + (\hat{\bf n}'\cdot{\bf B}({\bf r}'))\nabla'G
                                       \right].
\label{KirchB}
\earr
Here $G$ denotes the Green's function of the scalar Helmholtz equation in free
space, thus $G$  solves
\be
[\nabla^2 + k^2] G({\bf r},{\bf r}') = - \delta^{(3)}({\bf r}-{\bf r}'),
\ee
where $k=\omega/c$,
and can be written 
\be
 G ({\bf r}, {\bf r}')= \frac {e^{ik|{\bf r}-{\bf r}'|}} { 4\pi |{\bf r}-{\bf r}'| }.
\ee
To get the total field at a point in free space, the integrals in 
Eqs.\ (\ref{KirchE}) and (\ref{KirchB}) 
should be over the surfaces of all the scatterers that are present, and 
the fields entering the integrals should be the exact fields 
at those interfaces.
Here we restrict the attention to the fields scattered from the sphere.
In that case, only the Hankel function terms in Eqs.\ (\ref{Eseries})
and (\ref{Bseries}) contribute. 
Evaluating the surface integrals 
using an expansion of the Green's function in terms of plane waves
(propagating and evanescent),
\be
 G({\bf r},{\bf r}')
 =i
 \int \frac{d^2q_\|}{(2\pi)^2}
 \frac{e^{i\sqrt{k^2-q_{\|}^2}|z-z'|}}{2\sqrt{k^2-q_{\|}^2}}
 e^{i{\bf q}_{\|}\cdot({\bf r}_{\|}-{\bf r}'_{\|})},
\label{Gcart}
\ee
one finds that the field outside the sphere
can be written
\barr
 &&  {\bf E}=
 \sum_{lm,\sigma}
 b_{lm}^{\sigma} 
 \int \frac{d^2 q_{\|}}{(2\pi)^2}
\breakeq
\times
 \left[
 g_{lm}^{s\sigma}({\bf q}) (\hat{\bf z}\crossprod \hat{\bf q}_{\|})
 - g_{lm}^{p\sigma}({\bf q})
  (\hat{\bf q} \crossprod (\hat{\bf z}\crossprod\hat{\bf q}_{\|}))
 \right]
 e^{i{\bf q}\cdot {\bf r}}.
\earr
Here $\sigma$ denotes a polarization [(E) or (M)]. 
The coupling factors are found to be
\be
 g_{lm}^{pE}({\bf q}) = 
 g_{lm}^{sM}({\bf q}) = 
 \frac{(-1)^{l+m+1} e^{-i{\bf q}\cdot{\bf R}_{\rm sph}}} 
         { 2 \sqrt{ k^2 - |{\bf q}_{\|}|^2 } }
   U_{l,-m}({\bf q}),
\ee
and
\be
 g_{lm}^{sE}({\bf q}) = 
  -\, \, g_{lm}^{pM}({\bf q}) = 
 \frac{(-1)^{l+m+1}\, e^{-i{\bf q}\cdot{\bf R}_{\rm sph}}} 
 { 2 \sqrt{ k^2 - |{\bf q}_{\|}|^2 } }
 V_{l,-m}({\bf q}),
\ee
where ${\bf R}_{\rm sph}$ is a vector pointing to the center of the sphere.
As for the $\phi_q$ dependence of $g_{lm}$, it is in view of 
Eq.\ (\ref{UVang}) clear that
$$
  g_{lm}^{\sigma\sigma'} \propto e^{i m \phi_q}.
$$

\end{multicols}

\newpage
\narrowtext
\begin{table}
\caption{
The off-diagonal element $\epsilon_1(\omega)$ of the dielectric tensor 
for Fe and Co for a few photon energies. The data are taken from 
Ref.\ \protect\onlinecite{stearns}.
}
\vspace{0.2 cm}
\begin{tabular}{dll}
$\hbar\omega$ (eV) & Fe & Co \\
\tableline
1.5 & --1.42--0.089i & --1.16--0.089i \\
2.0 & --0.67--0.26i & --0.40--0.15i  \\
2.5 & --0.26--0.22i & --0.20--0.096i \\
3.0 & --0.13--0.12i & --0.10--0.067i \\
\end{tabular}
\label{eps1tab}
\end{table}


\begin{references}
\bibitem{Nonlocalnote}
We use a local dielectric function.  Nonlocal effects are small and would 
not change the results of this calulation in any considerable way.
For a more detailed discussion of this, see 
Sec.\ IV D of Ref.\ \protect\onlinecite{PJ98}.
\bibitem{Notesecond}
Note that strictly speaking there are
contributions in Eqs.\ (\protect\ref{Erads}) and (\protect\ref{Eradp})
that are second order 
in $\epsilon_1$. They are, however, negligibly small.
\end{references}
\end{document}